\documentclass[12pt,twoside,a4paper]{irs}

\newcommand {\vsp}   {\vspace*}

\def\title#1{\vsp{-16mm}\begin{center}\Large\bf{#1}\end{center}\vsp{0mm}}
\def\author#1{{\begin{center}\textbf{#1}\end{center}\vspace{-1mm}}}
\def\address#1{\vsp{-3mm}\begin{center}\baselineskip12pt\normalsize{#1}\end{center}\vsp{-1mm}}

\def\abstract#1{{\vspace{-5mm}
    \begin{center}
      \begin{minipage}{0.85\textwidth}
        \noindent\bf \textit{Abstract:}
        \small\rm\emph{#1}
				\vsp{-0.5em}
      \end{minipage}
    \end{center}
}}

\def\authorsheadline=#1{\global\def\@authorsheadline{#1}}
\global\def\@authorsheadline{}

\def\TeX{T\kern-.1667em\lower.5ex\hbox{E}\kern-.125emX}

\def\LaTeXG{{\rm L\kern-.36em\raise.3ex\hbox{\sc a}\kern-.15emT\kern-.1667em\lower.7ex\hbox{E}\kern-.125emX}}
\def\LaTeXK{{\it L\kern-.24em\raise.4ex\hbox{\scriptsize \it A}\kern-.20emT\kern-.1667em\lower.5ex\hbox{E}\kern-.125emX}}

\usepackage{svg}
\usepackage{epsf,exscale,times}
\usepackage[english]{babel}
\usepackage{graphicx}
\usepackage{amsmath}
\usepackage{fancyhdr}
\usepackage{svg}
\usepackage{siunitx}
\usepackage{cite}
\usepackage{amsmath,amssymb,amsfonts}

\usepackage{graphicx}
\usepackage{textcomp}
\usepackage{xcolor}
\usepackage{subcaption}
\usepackage{algorithm}
\usepackage{algpseudocode}
\usepackage{float}

\usepackage[acronym]{glossaries}
\usepackage{hyperref}
\usepackage{ifthen}
\usepackage{textgreek}
\usepackage{ifthen}
\usepackage{gensymb}
\usepackage{textcomp}

\newacronym{RCS}{RCS}{radar cross section}
\newacronym{MoM}{MoM}{Method of Moments}
\newacronym{FMM}{FMM}{Fast Multipole Method}
\newacronym{RL-GO}{RL-GO}{Ray-launching Geometrical Optics}
\newacronym{PEC}{PEC}{perfect electric conductor}
\newacronym{CS}{CS}{chirp sequence}
\newacronym{FMCW}{FMCW}{frequency-modulated continuous wave}
\newacronym{FFT}{FFT}{fast Fourier transform}
\newacronym{SNR}{SNR}{signal-to-noise ratio}
\newacronym{SBR}{SBR}{shooting and bouncing rays}
\newacronym{GPU}{GPU}{Graphics Processing Unit}
\newacronym{AoA}{AoA}{angle of arrival}
\newacronym{AWGN}{AWGN}{additive white Gaussian noise}
\newacronym{UWB}{UWB}{ultra wideband}
\newacronym{AMR}{AMR}{autonomous mobile robot}
\newacronym{LOS}{LOS}{line of sight}
\newacronym{LRP}{LRP}{Local Reference Point}
\newacronym{PSO}{PSO}{Particle Swarm Optimization}
\newacronym{RMSE}{RMSE}{root-mean-square error}
\usepackage{microtype}

\usepackage[font=small,skip=6.5pt]{caption}

\begin{document}

%%----------------------------------------------
%% Copyright notice (c) 2025 DGON
%%----------------------------------------------
\fancypagestyle{firststyle}
{
   \fancyhf{}
   \lfoot{ \footnotesize{The International Radar Symposium IRS 2025, May 21-23, 2025, Hamburg\break     ISBN 978-3-00-079305-9 \copyright  2024 DGON} }
   \rfoot{ \footnotesize {\thepage} }
}

\thispagestyle{firststyle}
\fancyhf{}
\renewcommand{\headrulewidth}{0pt}
\renewcommand{\footrulewidth}{1pt}
\renewcommand{\footskip}{50pt}

\pagestyle{fancy}
\fancyfoot[RO,LE]{ \footnotesize {\thepage} }

%%----------------------------------------------------
%% Insert the title of your paper between the brackets
%%----------------------------------------------------
\title{Hybrid baseband simulation for single-channel radar-based indoor localization system}

%%---------------------------------------------
%% Insert authors name(s) and address(es)
%% mark the presenting author with $^\ast$
%%---------------------------------------------

\author{
Sven Hinderer$^{*}$, Zheming Yin$^{*}$, Athanasios Papanikolaou$^{**}$, Jan Hesselbarth$^{**}$, Bin Yang$^{*}$}
\address{
    $^{*}$Institute of Signal Processing and System Theory, University of Stuttgart, Stuttgart, Germany\\
    $^{**}$Institute of Radio Frequency Technology, University of Stuttgart, Stuttgart, Germany\\
		email: surname.lastname@\{iss, ihf\}.uni-stuttgart.de
		}

\noindent\textbf{Note:} This is the accepted version of the paper. 
The final version is published in the \emph{Proceedings of the IEEE 2025 26th International Radar Symposium (IRS)}. 
DOI: \href{https://doi.org/10.23919/IRS64527.2025.11046115}{10.23919/IRS64527.2025.11046115}

\vspace{1em}

\renewcommand{\thefootnote}{}
\footnotetext{
	\textcopyright~2025 Personal use of this material is permitted.  Permission from IEEE must be obtained for all other uses, in any current or future media, including reprinting/republishing this material for advertising or promotional purposes, creating new collective works, for resale or redistribution to servers or lists, or reuse of any copyrighted component of this work in other works
}
\renewcommand{\thefootnote}{\arabic{footnote}}

%%------------------------------------
%% The abstract starts here
%%------------------------------------
\abstract{
Indoor localization with chirp sequence radar at millimeter wavelength offers high localization accuracy at low system cost. We propose a hybrid radar baseband signal simulator for our novel single-channel radar-based indoor localization system consisting of an active radar and passive reflectors as references. By combining ray tracing channel simulations with real measurements of the two-way antenna gain of the radar and accurate simulation of the radar cross section of chosen reflectors, realistic modeling of the baseband receive signal in complex scenarios is achieved.
}

%%------------------------------------
%% The introduction starts here
%%------------------------------------

\section{Introduction}
\begin{figure}[h!]
	\centering
	\begin{subfigure}[t]{0.44\textwidth}
	\includegraphics[width=\textwidth]{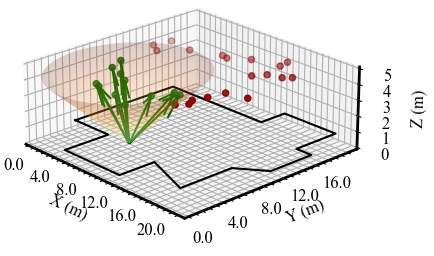}	
	\caption{Our conceptual system. A radar (at the cone vertex) senses LRPs (dots, green dots are detected) above it and uses them for localization~\cite{ipin_sven}.}
	\label{fig:cone}
	\end{subfigure}
	\qquad
	\begin{subfigure}[t]{0.48\textwidth}
	\includesvg[inkscapelatex=false,width=\textwidth]{lab.svg}	
	\caption{The demonstrator system in development including four octahedral corner reflectors and a small~\gls{AMR} that will be equipped with the radar.}
	\label{fig:demo}
	\end{subfigure}

	\caption{Depiction of a) our high-level localization system simulator - well suited for LRP placement optimization \cite{ipin_sven}, but not for realistic system evaluation and b) the actively developed demonstrator - provides real sensor signals, but is limited to the laboratory and time-consuming to set up.}
	\label{fig:two-way_measurement}
\end{figure}

One solution for indoor localization with decimeter and centimeter  accuracy as required for many \gls{AMR} operations is \gls{UWB} technology \cite{uwb}. 
We previously introduced a novel single-channel radar-based \gls{UWB} indoor localization system \cite{ipin_pascal, ipin_sven} that offers high accuracy at low system cost. An active radar placed on a moving \gls{AMR} faces the ceiling. Passive radar reflectors are distributed on the ceiling at fixed and known positions and serve as passive (low-cost) \glspl{LRP} for \gls{AMR} localization. The conceptual system is depicted in Fig.~\ref{fig:cone} and an image of the demonstrator system currently in development is given in Fig.\ref{fig:demo}. In this work, we present a hybrid radar baseband simulator to bridge the gap between Fig.~\ref{fig:cone} and Fig.~\ref{fig:demo}.

Due to their low cost and commercial availability, we choose octahedral corner reflectors~\cite{octahedral} as shown in Fig.~\ref{fig:demo} as~\glspl{LRP}. Other suitable millimeter wave retroreflectors that are still researched could be 3D-printed Luneburg lenses~\cite{3D_luneburg} or Van-Atta arrays~\cite{van_atta_inkjet, van_atta_60GHz, millimetro, fmcwindoor}. Van-Atta arrays would allow cross-polarized reflection and (semi-)active modulation of the reflected signal for easier detection and identification of~\glspl{LRP}.

Since our simple, passive corner reflectors do not allow identification of~\glspl{LRP}, we face a global, feature-based localization problem with unknown correspondences~\cite{prob_robotics}. Our~\gls{LRP} features for localization are the range, Doppler and amplitude of detected~\glspl{LRP}. The correspondances, i.e. the IDs of~\glspl{LRP}, can then be estimated through recursive, probabilistic \gls{AMR} tracking using a particle filter~\cite{ipin_pascal, ipin_sven}. In our previous work~\cite{ipin_sven}, we proposed an advanced~\gls{PSO} algorithm that optimizes the~\gls{LRP} placement such that the two extreme cases of localization with perfectly known~\gls{LRP} IDs and completely unknown~\gls{LRP} IDs are improved, hence increasing the localization performance and robustness of the system.

Accurate and computationally tractable channel simulations enable fast prototyping and testing of such systems. Ray tracing offers this combination and has been applied to simulate various millimeter wave radar systems. In \cite{bleder_raytrace_iss} Blender\textsuperscript{\textregistered} and its optical ray tracer have been used to simulate the \gls{FMCW} baseband signal for static indoor scenes. Blender\textsuperscript{\textregistered} has also been applied in \cite{blender_raytrace_dynamic} to simulate \gls{FMCW} radar signals in dynamic scenes for gesture recognition. The authors of \cite{nvidia_optics__raytrace_pulse_radar} developed a ray tracer for pulsed radar signals based on the NVIDIA OptiX\textsuperscript{\texttrademark} engine. The commercial software Wireless InSite\textsuperscript{\textregistered} was applied in  \cite{60ghz_WI_materials}, showing that precise material properties are key for accurate simulation results. Similar findings are given in \cite{radarays_raytracing_rotating_radar}, where a custom ray tracer is developed for indoor localization using rotating \gls{FMCW} radars. In \cite{fmcw_pedestrian_automotive_raytrace} another custom \gls{FMCW} ray tracer is developed to estimate the \gls{RCS} of pedestrians for automotive applications. Further automotive~\gls{FMCW} radar simulators are given in \cite{automotive_raytrace_mono, automotive_raytrace_array}, where radar frontend simulations are coupled with their own Matlab\textsuperscript{\textregistered} ray tracer for single- and multi-channel radar systems. 

In our work, we rely on the built-in ray tracer of Matlab\textsuperscript{\textregistered}~\cite{matlabraytracer} to simulate the radar signal reflections by the room and combine it with real measurements of the radar antennas and accurate simulations of the \gls{RCS} of our reflectors. Thereby, a realistic channel simulation with special focus on the \gls{LRP} reflectors used for localization is obtained. 
\section{Signal model}
\Gls{CS} radars are \gls{FMCW} radar systems with short frequency ramps. The Infineon BGT60TR13C~\cite{infineon2023bgt60tr13c_datasheet} used in our project and depicted in Fig.~\ref{fig:radar_mount} has one transmit and three colocated receive antennas. Each receiver has one (In-phase) channel. Compared to \gls{CS} radar systems with complex I/Q processing, this may cost up to $\SI{3}{dB}$ \gls{SNR}~\cite{real_complex_baseband_TI}. However, it circumvents the need for I/Q calibration and requires less hardware components. Our localization system uses only a single receiver. The elements of the baseband receive signal matrix $\mathbf{X}\in\mathbb{R}^{M\times N}$ per receiver are given by
\begin{align}
	x\left(m,n\right)
	\label{eq:baseband} = \sum_{k=1 }^{K}A_k\cos\left(-2\pi\left(\frac{2Br_k}{T_{chirp}c}T_m m+\frac{2f_0 v_k}{c}\left(T_m m+T_nn\right)+\phi_k\right)\right)+\nu(m,n).
\end{align}
The receive signal is a superposition of $K$ point target reflections with fast time samples $0\leq m\leq M-1$ and slow time samples $0\leq n\leq N-1$ used for range- and Doppler estimation. Each reflection $k$ has an associated amplitude $A_k$, range $r_k$, relative velocity $v_k$ w.r.t. the radar and phase $\phi_k$. $B$, $T_{chirp}$ and $f_0$ denote the bandwidth, duration and start frequency of the chirps, $c$  is the speed of light, $T_m$ and $T_n$ are the fast- and slow time sampling intervals and $\nu$ is \gls{AWGN}. Unknown signal interferences by the radar frontend and \gls{RCS} fluctuations are omitted for simulation. The above formula is derived under the assumptions of a linear channel, a constant radar velocity, the narrowband assumption $(B\ll f_0)$ and the relativistic assumption $(v_k\ll c)$. A detailed derivation can be found in~\cite{phd_gor}.

Our baseband signal simulation is split into two parts. The radar reflectors on the ceiling are used for localization and their directly reflected power is therefore modeled more accurately than with the ray tracer. $L$ of $K$ summands in \eqref{eq:baseband} corresponding to $L$ reflectors with \gls{LOS} to the radar are computed using the radar equation and simulated \gls{RCS} of the reflectors. The remaining $K-L$ summands are $K-L$ rays from the Matlab\textsuperscript{\textregistered} ray tracing channel simulation \cite{matlabraytracer} of the room without reflectors.

\subsection{Ray tracing channel simulation}
To simulate electromagnetic wave propagation, the \gls{SBR} method ~\cite{sbr} implemented in the Antenna Toolbox of Matlab\textsuperscript{\textregistered}~\cite{matlabraytracer} launches uniformly spaced rays from the transmitter, traces their propagation paths and collects all rays that fall into the reception sphere of the receiver. Its implementation allows for 3D simulation of multiple surface reflections and up to two edge diffractions for a frequency range from \SI{100}{MHz} to \SI{100}{GHz} including polarization. Refraction, corner diffraction, diffuse surface scattering and \gls{GPU} accelerated computations are currently not supported.

The ray tracer outputs the path loss $L_k$ in dB with a modified Friis equation, the propagation distance $d_k=2r_k$, the \gls{AoA} $AoA_k$ with azimuth and elevation and the phase $\phi_k$ for each collected ray corresponding to a point target reflection $k$. The voltage amplitude $A_k$ of the sinusoidal receive signal is given by $A_k=\sqrt{2P_T 10^{-L_k/10}R}$, where $P_T$ is the transmit power and $R=\SI{50}{Ohm}$ is the assumed impedance over which the voltage is measured. The relative velocity $v_k$ is computed by projecting the velocity vector of the radar onto $AoA_k$. Better Doppler estimates for rays with multiple reflections are described in~\cite{domus} but are omitted due to Matlab\textsuperscript{\textregistered} source code restrictions. Finally, the noise voltage is drawn from a Gaussian $\nu(m,n)\sim \mathcal{N}(0, \sigma_N^2)$ with noise variance $\sigma_N^2 = P_N R$. $P_N=k_BT_0FB_R$ is the receiver noise power \cite{richards_modern_radar} with Boltzmann's constant $k_B$, standard room temperature $T_0=\SI{290}{K}$, receiver noise figure $F$ (in linear scale) and instantaneous receiver bandwidth $B_R=1/(2T_m)$.
\subsection{Reflector target simulation}
The range $r_k$ and relative velocity $v_k$ between the radar and the center of the \gls{LOS} reflectors are computed as with the ray tracer. The phase is $\phi_k=2r_k\lambda$~\cite{phd_gor}, where $\lambda = c/f_0$ is the wavelength. $A_k$ is derived from the path loss with the radar equation $L_k=-10 \log_{10}\left(\frac{G_T G_R\lambda^2\sigma}{\left(4\pi\right)^3 r_k^4}\right)$~\cite{richards_modern_radar} with $\sigma$ being the mean \gls{RCS} of the reflectors and $G_T$ and $G_R$ denoting the gain of the transmit and receive antennas. As shown in the following section, the co-polarizations are almost identical $\sigma_{VV} \approx \sigma_{HH}$ 
and cross-polarization is negligible $\sigma_{VH} \approx \sigma_{HV} \approx 0$, which allows for the simplification of a scalar \gls{RCS}~$\sigma$ and omission of the polarization loss $l\approx1$ at the receiver.

\subsection{Radar cross section simulation of octahedral corner reflectors}
\begin{figure}[h!]
	\centering
	\begin{subfigure}[t]{0.23\textwidth}
		\includegraphics[width=\textwidth]{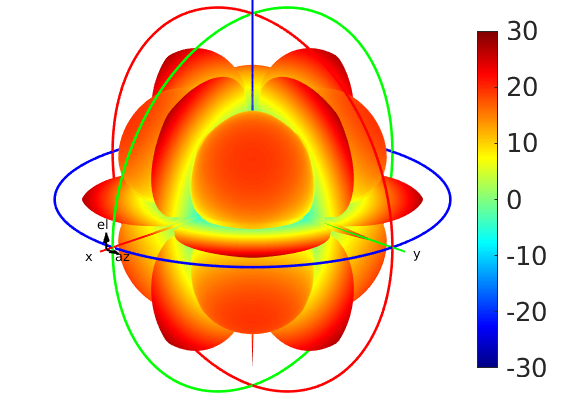}
		\caption{$\sigma_{VV}$}
		\label{fig:2GHz_VV}
	\end{subfigure}
	\hfill
	\begin{subfigure}[t]{0.23\textwidth}
		\includegraphics[width=\textwidth]{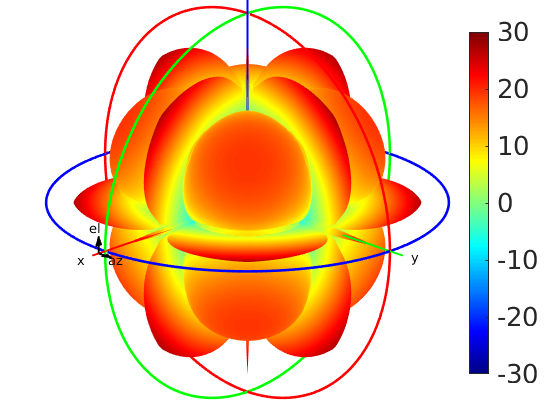}
		\caption{$\sigma_{HH}$}
		\label{fig:2GHz_HH}
	\end{subfigure}
	\hfill
	\begin{subfigure}[t]{0.23\textwidth}
		\includegraphics[width=\textwidth, angle= 0]{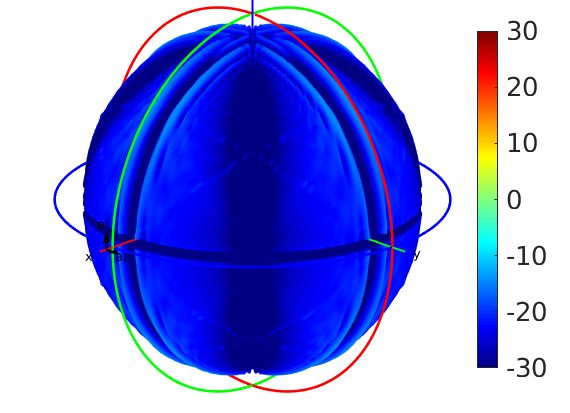}
		\caption{$\sigma_{VH}$}
		\label{fig:2GHz_VH}
	\end{subfigure}
	\hfill
	\begin{subfigure}[t]{0.23\textwidth}
		\includegraphics[width=\textwidth, angle= 0]{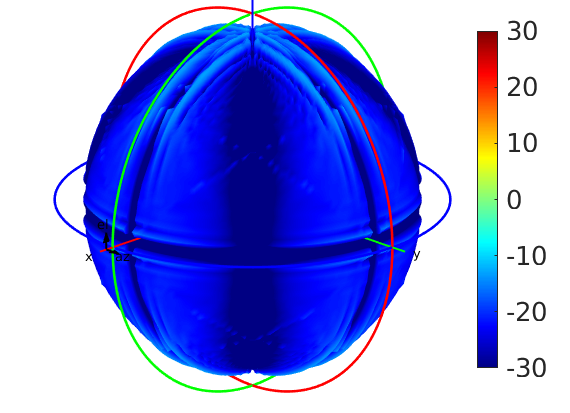}
		\caption{$\sigma_{HV}$}
	\label{fig:2GHz_HV}
	\end{subfigure}

	\caption{Feko\textsuperscript{\textregistered} \gls{RL-GO} \gls{RCS} simulation results in dBsm for $f_0=\SI{59}{GHz}$ and $B=\SI{2}{GHz}$.}
	\label{fig:rcs}
\end{figure}

As the Matlab\textsuperscript{\textregistered} ray tracer is not capable of correct simulation of the crucial \gls{RCS} $\sigma$ of the radar reflectors used as position references for localization, we resort to more accurate simulation techniques. The large electric size of the chosen octahedral corner reflectors from Fig.~\ref{fig:demo} with square $\left(21.5, 21.5, 0.1\right)\,\mathrm{cm}$ aluminum plates prohibits the application of exact numerical methods like \gls{MoM} and \gls{FMM}. Thus, Feko\textsuperscript{\textregistered} and its asymptotic \gls{RL-GO} solver is used. Feko\textsuperscript{\textregistered}'s \gls{RL-GO} gives accurate \gls{RCS} simulation results for electrically large \glspl{PEC} like corner reflectors~\cite{feko_rlgo}.
To estimate the \gls{RCS} for a given bandwidth, the~\gls{RCS} at the start and end frequencies of the chirp and the resulting patterns in linear scale $\sigma_{VV}=\frac{1}{2}\left(\sigma_{VV}^{f_0} + \sigma_{VV}^{f_0+B}\right)$ and $\sigma_{HH}=\frac{1}{2}\left(\sigma_{HH}^{f_0} + \sigma_{HH}^{f_0+B}\right)$ are averaged as $\sigma=\frac{1}{2}\left(\sigma_{VV}+\sigma_{HH}\right)$. A simulation example is given in Fig.~\ref{fig:rcs}. 
\vspace{-1.0em}
\subsection{Two-way antenna gain measurement}
Our baseband simulation considers the normalized gain of the radar antennas, which rotate according to the robot's heading. Since the antennas of the BGT60TR13C are integrated with the radar chip, directly accessing the antenna ports and therefore standard antenna radiation pattern measurements are not possible. The current approach to compute the antenna radiation pattern for such integrated systems is a two-way antenna gain measurement $G^2=G_TG_R$ against a strong radar reflector. Similar measurements have previously been performed for 2D two-way antenna gains by Infineon~\cite{infineon2023bgt60tr13c_shield} and are also found in the literature~\cite{two_way_first, two_way_real}. Different to these experiments, we perform 3D radiation pattern measurements as required for localization. We further carry out measurements for all 3 receive antennas  with four different chirp bandwidths of $\left\{0.5, 1, 2, 4\right\}\,\mathrm{GHz}$ with all chirps starting at $f_0=\SI{59}{GHz}$. Thus, 12 two-way antenna gains are measured for simulation with different system settings.

\begin{figure}[h!]
	\centering
	\begin{subfigure}[t]{0.32\textwidth}
		\includegraphics[width=\textwidth]{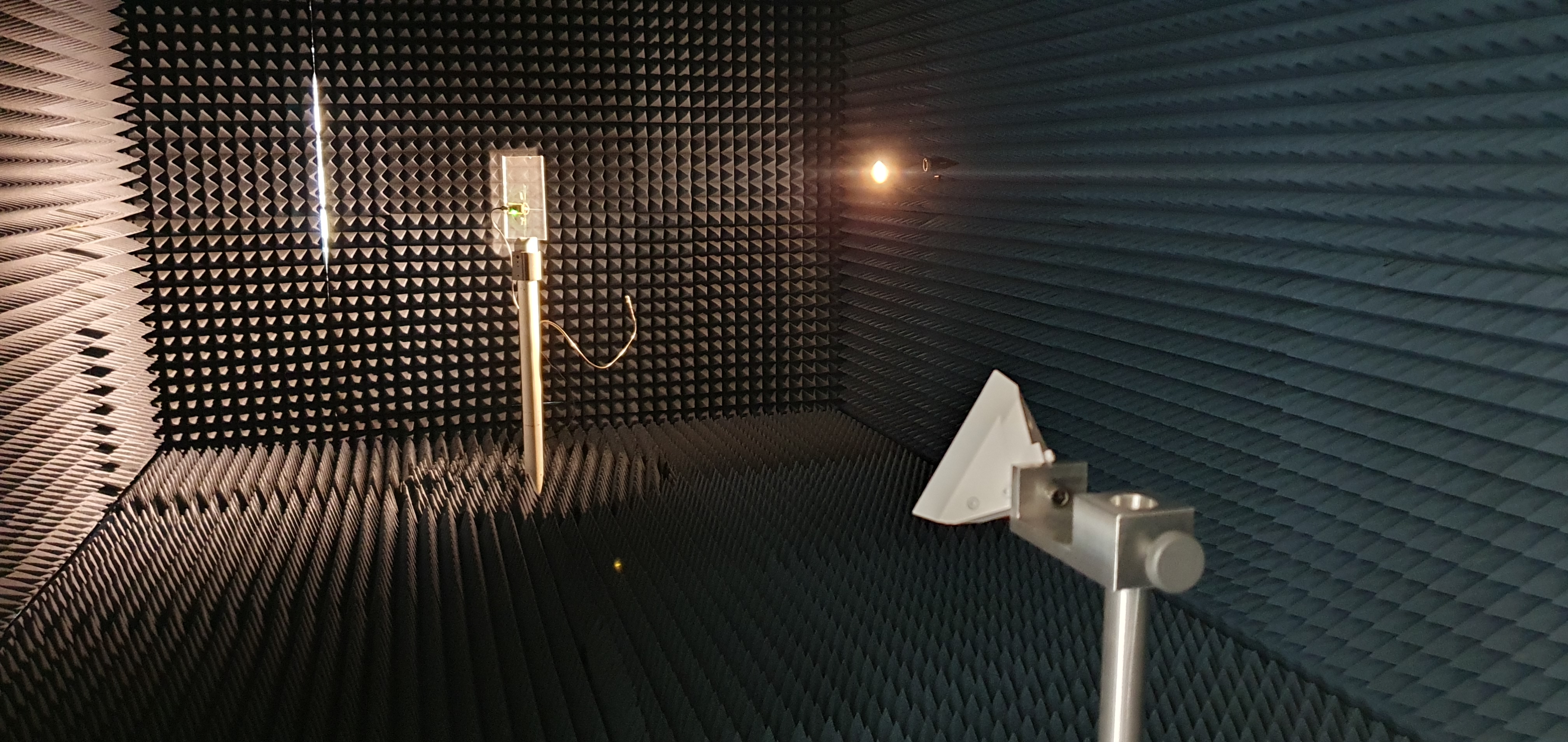}
		\caption{View from the reflector.}
		\label{fig:sicht_reflektor}
	\end{subfigure}
	\hfill
	\begin{subfigure}[t]{0.32\textwidth}
		\includesvg[inkscapelatex=false, width=\textwidth]{sicht_reflektor.svg}
		\caption{View from the radar and visualization of both rotation axes.}
		\label{fig:sicht_radar}
	\end{subfigure}
	\hfill
	\begin{subfigure}[t]{0.32\textwidth}
		\includesvg[inkscapelatex=false,width=\textwidth, angle= 0]{radar_mount2}
		\caption{Radar mount with manual angle adjustment.}
		\label{fig:radar_mount}
	\end{subfigure}
	
	\caption{Depiction of the two-way antenna gain measurement setup. The radar in c) is rotated with the green acrylic angle template by hand and with the rotary table visible in a) and b). We measure against the trihedral corner reflector shown in a) and b).}
	\label{fig:two-way_measurement}
\end{figure}

\begin{figure}[h!]
	\centering
	\begin{subfigure}[t]{0.48\textwidth}
		\includesvg[inkscapelatex=false, width=\textwidth]{range_profile}
		\caption{Amplitude extraction in range profile.}
		\label{fig:range_profile2.1}
	\end{subfigure}
	\qquad
	\begin{subfigure}[t]{0.45\textwidth}
		\includegraphics[width=\textwidth]{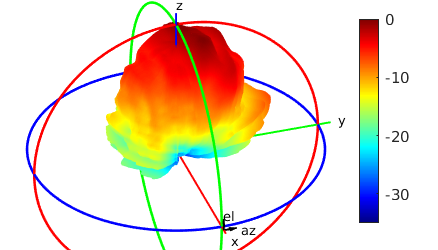}
		\caption{Normalized two-way antenna gain in dB.}
		\label{fig:range_profile2.2}
	\end{subfigure}
	
	\caption{Two-way radiation pattern measurement example for Rx2 and $\SI{2}{GHz}$ bandwidth: a) averaged range profile $\underline{\bar{R}}$ (for one angle) and b) normalized 3D two-way antenna gain $G^2=G_{T}G_{R2}$.}
	\label{fig:rangeprofile2}
\end{figure}

Our measurement setup is shown in Fig.~\ref{fig:two-way_measurement}. We use an anechoic chamber of the Institute of Radio Frequency Technology for millimeter wave measurements from $\SI{26.5}{GHz}$ to $\SI{110}{GHz}$.
A trihedral corner reflector visible in Fig.~\ref{fig:sicht_reflektor} and Fig.~\ref{fig:sicht_radar} is chosen as reference. This type of reflector offers a large \gls{RCS} while being polarization independent and small enough to satisfy the far-field assumption in the anechoic chamber. Fig.~\ref{fig:radar_mount} shows how the radar is mounted on top of an acrylic angle template and an additional acrylic plate for the azimuth adjustment of the radar by hand. The elevation is set with the rotary table that the radar is placed on, as seen in Fig.~\ref{fig:sicht_radar}.
The two-way antenna gain per angle can be computed by measuring the receive power that is reflected by the trihedral. Therefore, we extract the peak amplitude of the reflector in the range profile of the receive signal. The range profile vector $\underline{R}\in\mathbb{R}^{M/2+1}$ for each chirp $n$ is computed by range~\gls{FFT} \cite{new_chirp_sequence} over the measured, real-valued baseband signal $x(m,n)$ with Hermitian symmetric spectrum and normalized such that the signal energy is preserved in the \gls{FFT} processing:
\begin{align}
 	R(o,n) &= \frac{|S(o,n)|^2}{M}\quad\mathrm{for}\quad o\in\left\{0,\frac{M}{2}\right\}\quad \mathrm{and}\quad R(o,n)=\frac{2|S(o,n)|^2}{M}\,\mathrm{else},\,\mathrm{with} \\
 	S(o,n) &= \sum_{m=0}^{M-1} x(m,n) e^{-j 2\pi \frac{om}{M}}, \quad 0\leq o\leq \frac{M}{2}.
	\label{eq:range_FFT}
\end{align}
For this antenna gain measurement, we choose a flattop window as range \gls{FFT} window due to their low amplitude error~\cite{dft}. For de-noising, $N=100$ range profiles from consecutive chirps are averaged, thus giving the de-noised range profile elements $\bar{R}(o)=\frac{1}{100}\sum_{n=0}^{99}R(o,n)$. Fig.~\ref{fig:range_profile2.1} depicts a resulting range profile after conversion of the frequency bins to their corresponding ranges. The mount of the trihedral has shown to reflect between 20 to 200 times less power than the reflector and was therefore ignored.
After extracting the peak power, the radar is moved to the next angle and the same procedure is repeated. Measurements are carried out with an angular resolution of $2\degree$ for both azimuth and elevation. Fig.~\ref{fig:range_profile2.2} gives a resulting radiation pattern. It can be inferred that sharp 3D patterns are collected with the described method. We are also able to reproduce the skewed mainlobe and the sidelobe in the E-plane of the antennas (right in Fig.~\ref{fig:range_profile2.2}), as shown in the 2D patterns measured by Infineon \cite{infineon2023bgt60tr13c_shield}.

\section{Results}
\begin{figure}[tbp]
	\centering
	\begin{subfigure}[t]{0.23\textwidth}
		\includegraphics[width=\textwidth]{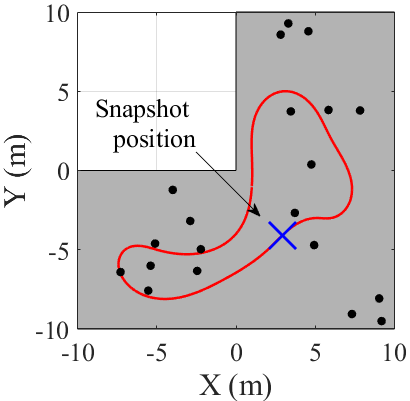}
		\caption{Top-down view of the room including the \gls{AMR} trajectory and snapshot position.}
		\label{fig:trajecory}
	\end{subfigure}
	\hfill
	\begin{subfigure}[t]{0.332\textwidth}
		\includegraphics[width=\textwidth]{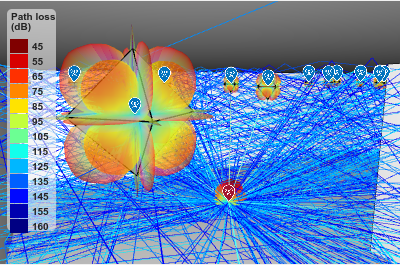}
		\caption{Rays from ray tracing and reflector target simulation in L-shaped room.}
		\label{fig:raytracing}
	\end{subfigure}
	\hfill
	\begin{subfigure}[t]{0.394\textwidth}
		\includegraphics[width=\textwidth]{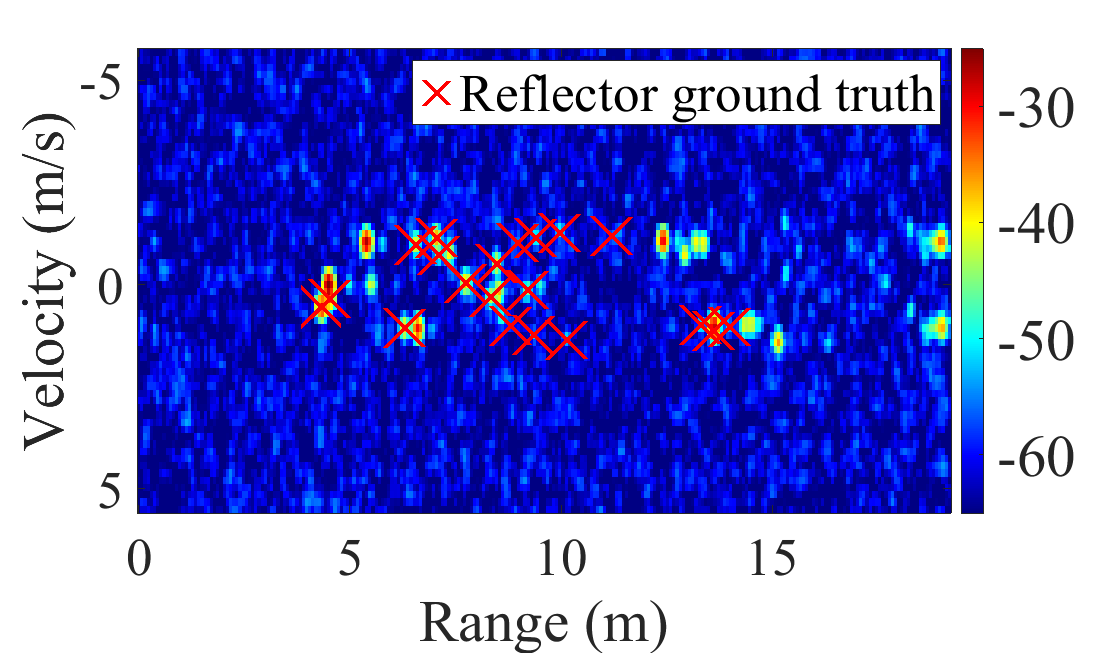}
		\caption{Range Doppler map in dB.}
		\label{fig:range_Doppler}
	\end{subfigure}
	\caption{Demonstration of the simulation results for one snapshot: a) \gls{AMR} trajectory in the room and positions of LRPs (black dots), b) visualization of all rays, the radar (red marker) and the reflectors (blue markers) and c) the corresponding range Doppler map.}
\label{fig:results}
\end{figure}

\begin{figure}[tbp]
	\centering
	\begin{subfigure}[t]{0.32\textwidth}
		\includegraphics[width=\textwidth]{range_Doppler0}
		\caption{$t_0$}
		\label{fig:t0}
	\end{subfigure}
	\hfill
	\begin{subfigure}[t]{0.32\textwidth}
		\includegraphics[width=\textwidth]{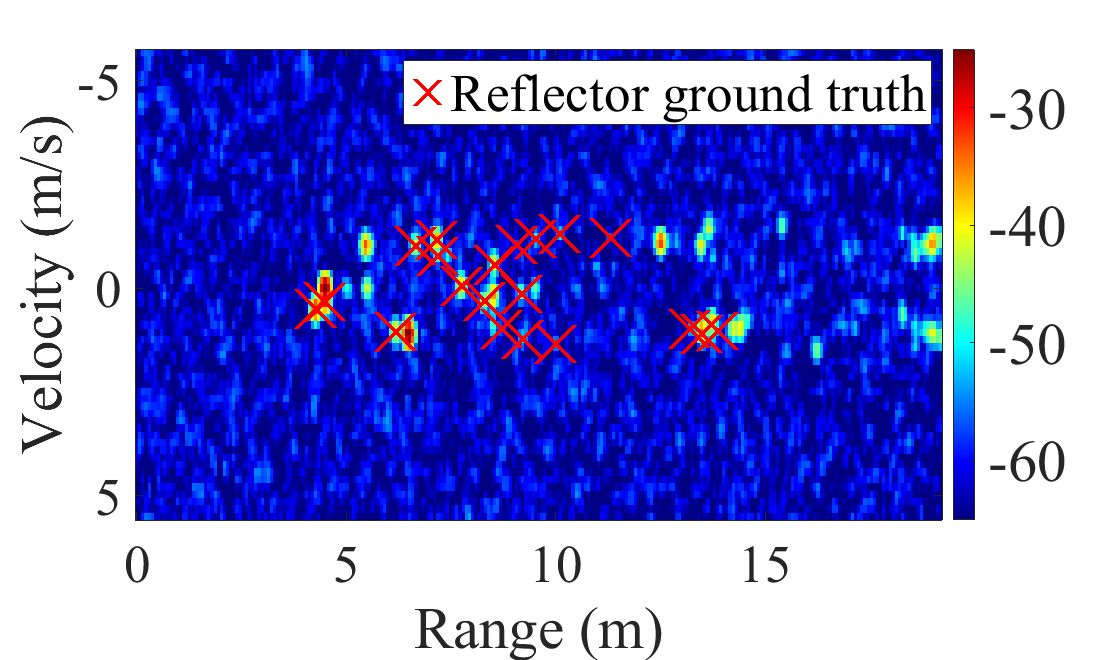}
		\caption{$t_1=t_0+100\,\mathrm{ms}$.}
		\label{fig:t1}
	\end{subfigure}
	\hfill
	\begin{subfigure}[t]{0.32\textwidth}
		\includegraphics[width=\textwidth, angle= 0]{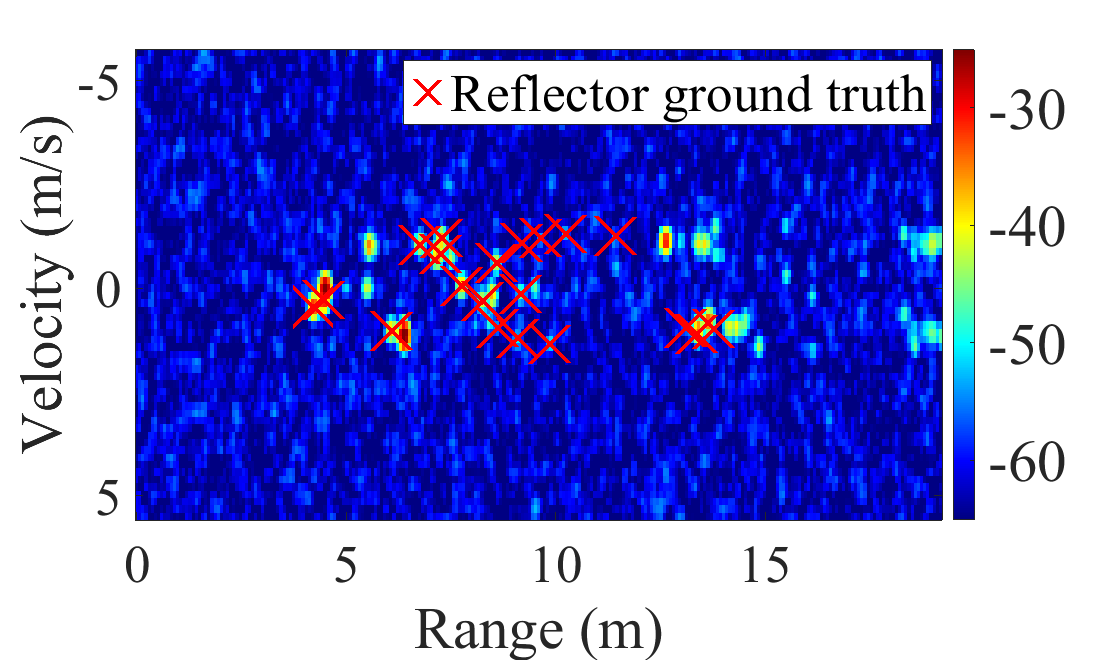}
		\caption{$t_2=t_0+200\,\mathrm{ms}$.}
		\label{fig:t2}
	\end{subfigure}
	\caption{Consecutive range Doppler maps in dB collected at different times $t_0, t_1, t_2$ along the trajectory show high transient consistency and realism of the simulated baseband signals.}
	\label{fig:transient}
\end{figure}

To evaluate our simulator, we use Blender\textsuperscript{\textregistered} to create a non-convex, L-shaped room of size $\left(20, 20, 5\right)\,\mathrm{m}$, where the radar at a height of $z=0.5\,\mathrm{m}$ does not have \gls{LOS} to all reflectors from every position and import it to Matlab\textsuperscript{\textregistered}. The walls are modeled as concrete, a commonly used material for buildings. Our \gls{AMR} moves along a realistic trajectory shown in Fig.~\ref{fig:trajecory} and we showcase the simulation results for one position in Fig.~\ref{fig:raytracing}.
For the verification of our simulation results, we compute the baseband signal $\mathbf{X}$ from \eqref{eq:baseband} for the snapshot in Fig.~\ref{fig:raytracing} and provide the resulting range Doppler map matrix $\mathbf{D}\in\mathbb{R}^{M/2+1 \times N}$ in Fig.~\ref{fig:range_Doppler}. The range Doppler map elements $D(o,p)$ are computed by Doppler~\gls{FFT}~\cite{new_chirp_sequence} over $S(o,n)$ from:

\begin{align}
	D(o,p) &= \frac{|T(o,p)|^2}{NM}\quad\mathrm{for}\quad o\in\left\{0,\frac{M}{2}\right\}\quad \mathrm{and}\quad D(o,p)=\frac{2|T(o,p)|^2}{NM}\,\mathrm{else},\,\mathrm{with} \\
	T(o,p) &= \sum_{n=0}^{N-1} S(o,n) e^{-j 2\pi \frac{pn}{N}}, \quad 0\leq p \leq N-1.
	\label{eq:Doppler_FFT}
\end{align}
For localization, we are more concerned with accurate estimation of the more reliable range and Doppler values than with the measured amplitudes of~\glspl{LRP}. Consequently, we apply Hamming windows for both range and Doppler \glspl{FFT} when computing range Doppler maps, as Hamming windows offer good frequency resolution~\cite{dft}. Further, the range Doppler map is post-processed by a convolution with a $5\times5$ Gaussian blur filter to add previously unmodeled target spread through missing diffuse scattering. As clearly visible in Fig.~\ref{fig:range_Doppler}, the reflectors are found at the correct range and Doppler values. There also exist multiple other peaks as expected in a heavily cluttered indoor environment, e.g. at \SI{4.5}{m} distance and zero Doppler caused by the ceiling. Further, we deduct from Fig.~\ref{fig:transient} that simulation of multiple consecutive signals along the trajectory with each simulation taken \SI{100}{ms} apart gives highly consistent results.

\begin{figure}[tbp]
	\centering
	\begin{subfigure}[t]{0.43\textwidth}
		\includesvg[width=\textwidth]{particlefilter}
		\caption{Snapshot of the particle filter localization with zoom into colored (weighted) particles denoting their likelihoods.}
		\label{fig:pft0}
	\end{subfigure}
	\qquad
	\begin{subfigure}[t]{0.43\textwidth}
		\includegraphics[width=\textwidth]{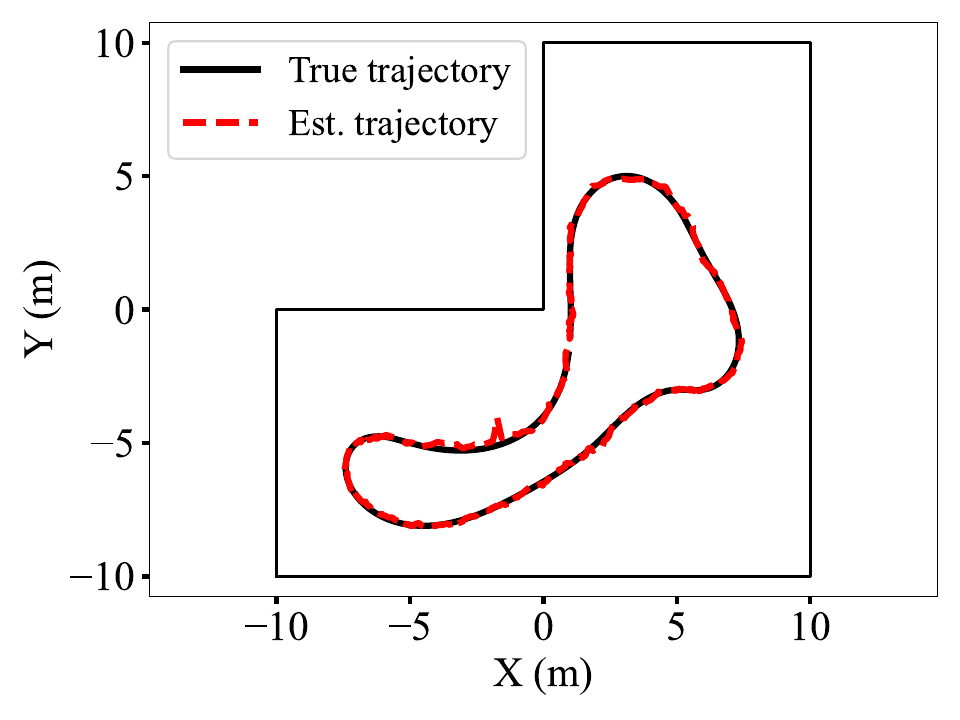}
		\caption{True and estimated \gls{AMR} trajectories with mean \gls{RMSE} of around \SI{13}{cm}.}
		\label{fig:pft1}
	\end{subfigure}
	\caption{Localization with a particle filter showing a) the particles, the \glspl{LRP} used for localization and the estimated \gls{AMR} position for the snapshot and b) the estimated \gls{AMR} trajectory.}
	\label{fig:particle_filter}
\end{figure}

The particle filter for localization, where particles form an empirical distribution over the space of \gls{AMR} poses, is depicted in Fig.~\ref{fig:particle_filter}. 
It is based on our previous simulator with a simplified, high-level detection model derived from Fig.~\ref{fig:cone}, not including any radar processing, false alarm, miss etc. \cite{ipin_sven}. The radar baseband simulator developed in this work now enables fast and realistic simulation of the whole localization system for rapid prototyping and optimization.
\section{Conclusion}
We have introduced a baseband signal simulator for our novel radar-based indoor localization system. To accurately simulate the reflections of the passive reflectors on the ceiling used as \glspl{LRP}, antenna measurements of the radar and \gls{RCS} simulations of the reflectors have been performed. These were combined with ray tracing simulations for channel modeling of the room.

Realistic simulation of our system could be achieved, which speeds up further development. In addition, the antenna measurements and \gls{RCS} simulations enable localization using not only range and Doppler, but also target amplitude information. 

\bibliographystyle{ieeetr}
\bibliography{ref}

\begin{thebibliography}{10}

\bibitem{uwb}
A.~Alarifi {\em et~al.}, ``Ultra wideband indoor positioning technologies:
  Analysis and recent advances,'' {\em Sensors}, vol.~16, no.~5, 2016.

\bibitem{ipin_pascal}
P.~Schlachter {\em et~al.}, ``Indoor positioning based on active radar sensing
  and passive reflectors: Concepts and initial results,'' in {\em Proceedings
  of the Work-in-Progress Papers at the 13th International Conference on Indoor
  Positioning and Indoor Navigation (IPIN-WiP 2023)}, pp.~1--16, 2023.

\bibitem{ipin_sven}
S.~Hinderer, P.~Schlachter, Z.~Yu, X.~Wu, and B.~Yang, ``Indoor positioning
  based on active radar sensing and passive reflectors: Reflector placement
  optimization,'' in {\em 2023 13th International Conference on Indoor
  Positioning and Indoor Navigation (IPIN)}, pp.~1--7, 2023.

\bibitem{octahedral}
A.~W. Doerry, ``Reflectors for {SAR} performance testing-second edition,''
  tech. rep., Sandia National Lab. (SNL-NM), Albuquerque, NM (United States),
  Feb 2014.

\bibitem{3D_luneburg}
R.~A. Bahr {\em et~al.}, ``{3D}-printed omnidirectional {Luneburg} lens
  retroreflectors for low-cost mm-wave positioning,'' {\em 2020 IEEE
  International Conference on RFID (RFID)}, pp.~1--7, 2020.

\bibitem{van_atta_inkjet}
J.~G.~D. Hester and M.~M. Tentzeris, ``Inkjet-printed flexible mm-wave
  {Van-Atta} reflectarrays: A solution for ultralong-range dense multitag and
  multisensing chipless {RFID} implementations for {IoT} smart skins,'' {\em
  IEEE Transactions on Microwave Theory and Techniques}, vol.~64, no.~12,
  pp.~4763--4773, 2016.

\bibitem{van_atta_60GHz}
K.~Trzebiatowski, M.~Rzymowski, L.~Kulas, and K.~Nyka, ``Simple millimeter wave
  identification system based on 60 {GHz Van Atta} arrays,'' {\em Sensors},
  vol.~22, no.~24, 2022.

\bibitem{millimetro}
E.~Soltanaghaei {\em et~al.}, ``Millimetro: Mmwave retro-reflective tags for
  accurate, long range localization,'' in {\em Proceedings of the 27th Annual
  International Conference on Mobile Computing and Networking}, 2021.

\bibitem{fmcwindoor}
L.~Marcaccioli {\em et~al.}, ``An accurate indoor ranging system based on
  {FMCW} radar,'' in {\em IEEE Intelligent Vehicles Symposium (IV)},
  pp.~981--986, 2011.

\bibitem{prob_robotics}
S.~Thrun, W.~Burgard, and D.~Fox, {\em Probabilistic robotics}.
\newblock Cambridge, Mass.: MIT Press, 2005.

\bibitem{bleder_raytrace_iss}
M.~Ouza, M.~Ulrich, and B.~Yang, ``A simple radar simulation tool for {3D}
  objects based on {Blender},'' in {\em 2017 18th International Radar Symposium
  (IRS)}, pp.~1--10, 2017.

\bibitem{blender_raytrace_dynamic}
A.~Ninos, J.~Hasch, M.~E.~P. Alvarez, and T.~Zwick, ``Synthetic radar dataset
  generator for macro-gesture recognition,'' {\em IEEE Access}, vol.~9,
  pp.~76576--76584, 2021.

\bibitem{nvidia_optics__raytrace_pulse_radar}
M.~Y. Martin {\em et~al.}, ``The design and implementation of a ray-tracing
  algorithm for signal-level pulsed radar simulation using the {NVIDIA
  OptiX\textsuperscript{\texttrademark} engine},'' {\em J. Commun.}, vol.~17,
  pp.~761--768, 2022.

\bibitem{60ghz_WI_materials}
H.~Obeidat {\em et~al.}, ``Channel impulse response at 60 {GHz} and impact of
  electrical parameters properties on ray tracing validations,'' {\em
  Electronics}, vol.~10, no.~4, 2021.

\bibitem{radarays_raytracing_rotating_radar}
A.~Mock, M.~Magnusson, and J.~Hertzberg, ``Radarays: Real-time simulation of
  rotating {FMCW} radar for mobile robotics via hardware-accelerated ray
  tracing,'' {\em IEEE Robotics and Automation Letters}, vol.~10,
  p.~2470–2477, Mar. 2025.

\bibitem{fmcw_pedestrian_automotive_raytrace}
Y.~Deep {\em et~al.}, ``Radar cross-sections of pedestrians at automotive radar
  frequencies using ray tracing and point scatterer modelling,'' {\em IET
  Radar, Sonar \& Navigation}, vol.~14, pp.~833--844, 2020.

\bibitem{automotive_raytrace_mono}
M.~Dudek {\em et~al.}, ``Millimeter wave {FMCW} radar system simulations
  including a {3D} ray tracing channel simulator,'' in {\em 2010 Asia-Pacific
  Microwave Conference}, pp.~1665--1668, 2010.

\bibitem{automotive_raytrace_array}
M.~Dudek {\em et~al.}, ``A versatile system simulation environment for
  millimeter-wave phased-array {FMCW}-radar sensors for automotive
  applications,'' in {\em Asia-Pacific Microwave Conference 2011},
  pp.~1478--1481, 2011.

\bibitem{matlabraytracer}
{The MathWorks Inc.}, {\em {Antenna Toolbox version: 9.4 (R2023b)}}, 2024.
\newblock Accessed: November 24, 2024. Available:
  \url{https://www.mathworks.com/}.

\bibitem{infineon2023bgt60tr13c_datasheet}
{Infineon Technologies AG}, {\em {BGT60TR13C Datasheet V2.4.9}}.
\newblock Munich, Germany, Nov. 2023.
\newblock Accessed: November 24, 2024. Available:
  \url{https://www.infineon.com/cms/en/product/sensor/radar-sensors/radar-sensors-for-iot/60ghz-radar/bgt60tr13c/}.

\bibitem{real_complex_baseband_TI}
K.~Ramasubramanian, ``Using a complex-baseband architecture in {FMCW} radar
  systems,'' tech. rep., Texas Instruments, May 2017.

\bibitem{phd_gor}
G.~Hakobyan, {\em Orthogonal Frequency Division Multiplexing Multiple-Input
  Multiple-Output Automotive Radar with Novel Signal Processing Algorithms}.
\newblock PhD thesis, Universit{\"a}t Stuttgart, Stuttgart, Feb 2018.

\bibitem{sbr}
H.~Ling {\em et~al.}, ``Shooting and bouncing rays: calculating the {RCS} of an
  arbitrarily shaped cavity,'' {\em IEEE Transactions on Antennas and
  Propagation}, vol.~37, no.~2, pp.~194--205, 1989.

\bibitem{domus}
A.~Battaglia and S.~Tanelli, ``Domus: Doppler multiple-scattering simulator,''
  {\em IEEE Transactions on Geoscience and Remote Sensing}, vol.~49, no.~1,
  pp.~442--450, 2011.

\bibitem{richards_modern_radar}
M.~A. Richards, J.~A. Scheer, and W.~A. Holm, {\em Principles of Modern Radar:
  Basic principles}.
\newblock The Institution of Engineering and Technology, 2010.

\bibitem{feko_rlgo}
A.~G. Aguilar {\em et~al.}, ``Overview of recent advances in the
  electromagnetic field solver {Feko},'' in {\em 2015 9th European Conference
  on Antennas and Propagation (EuCAP)}, pp.~1--5, 2015.

\bibitem{infineon2023bgt60tr13c_shield}
{Infineon Technologies AG}, {\em {BGT60TR13C Shield: XENSIV™ 60 GHz Radar
  System Platform}}.
\newblock Munich, Germany, Feb. 2023.
\newblock Application Note, Revision 2.40. Accessed: November 24, 2024.
  Available:
  \url{https://www.infineon.com/cms/en/product/sensor/radar-sensors/radar-sensors-for-iot/60ghz-radar/bgt60tr13c/}.

\bibitem{two_way_first}
L.~Piotrowsky {\em et~al.}, ``Antenna pattern characterization with an
  industrial robot assisted {FMCW} radar system,'' in {\em 2019 IEEE
  Asia-Pacific Microwave Conference (APMC)}, pp.~153--155, 2019.

\bibitem{two_way_real}
A.~C. Granich {\em et~al.}, ``Radiation pattern measurements using an active
  radar module,'' in {\em 2022 Antenna Measurement Techniques Association
  Symposium (AMTA)}, pp.~1--5, 2022.

\bibitem{new_chirp_sequence}
H.~Rohling and M.~Kronauge, ``New radar waveform based on a chirp sequence,''
  in {\em 2014 International Radar Conference}, pp.~1--4, 2014.

\bibitem{dft}
G.~Heinzel, A.~R{\"u}diger, and R.~Schilling, ``{Spectrum and spectral density
  estimation by the Discrete Fourier transform (DFT), including a comprehensive
  list of window functions and some new at-top windows}.'' 2002.

\end{thebibliography}

\end{document}